\documentclass[useAMS,usenatbib]{mn2e}
\usepackage{graphicx}
\usepackage{lscape}

\title[A new Y dwarf: WISE J0304-2705]{Discovery of a new Y dwarf: WISE J030449.03-270508.3}
\author[D. J. Pinfield et~al.]{D. J. Pinfield$^{1}$\thanks{E-mail: D.J.Pinfield@herts.ac.uk}, 
M. Gromadzki$^{2,3}$, S. K. Leggett$^{4}$, J. Gomes$^{1}$, N. Lodieu$^{5,6}$, 
\newauthor
R. Kurtev$^{3,2}$, A. C. Day-Jones$^{1}$, M. T. Ruiz$^{7}$, N. J. Cook$^{1}$,  C. V. Morley$^{8}$, 
\newauthor
M. S. Marley$^{9}$, F. Marocco$^{1}$, R. L. Smart$^{10}$, H. R. A. Jones$^{1}$, P. W. Lucas$^{1}$, 
\newauthor
Y. Beletsky$^{11}$, V. D. Ivanov$^{12}$, B. Burningham$^{1}$, J. S. Jenkins$^{7}$, C.Cardoso$^{10}$, 
\newauthor
J. Frith$^{1}$, J. R. A. Clarke$^{2}$, M. C. G\'{a}lvez-Ortiz$^{13}$, Z. Zhang$^{5,6}$
\\
$^{1}$Centre for Astrophysics Research, Science and Technology Research Institute, University of Hertfordshire, Hatfield AL10 9AB, UK \\
$^{2}$The Millennium Institute of Astrophysics (MAS), Santiago, Chile \\
$^{3}$Instituto de F\'{i}sica y Astronom\'{i}a, Universidad de Valpara\'{i}so, Av. Gran Breta\~{n}a 1111, Playa Ancha, Casilla 5030, Chile \\
$^{4}$Gemini Observatory, Northern Operations Center, 670 N. A'ohoku Place, Hilo, HI 96720, USA \\
$^{5}$Instituto de Astrof\'isica de Canarias (IAC), Calle V\'ia L\'actea s/n, E-38200 La Laguna, Tenerife, Spain \\
$^{6}$Departamento de Astrof\'isica, Universidad de La Laguna (ULL), E-38206 La Laguna, Tenerife, Spain \\
$^{7}$Universidad de Chile, Santiago, Casilla 36-D, Chile \\
$^{8}$Department of Astronomy and Astrophysics, University of California, Santa Cruz \\
$^{9}$NASA Ames Research Center \\
$^{10}$Istituto Nazionale di Astrofisica, Osservatorio Astronomico di Torino, Strada Osservatrio 20, 10025 Pino Torinese, Italy \\
$^{11}$Las Campanas Observatory, Carnegie Institution of Washington, Colina el Pino, 601 Casilla, La Serena, Chile \\
$^{12}$ESO, Av. Alonso de Cordova 3107, 19001 Casilla, Santiago 19, Chile \\
$^{13}$Centro de Astrobiolog\'ia (CSIC-INTA), Ctra. Ajalvir km 4, E-28850 Torrej\'on de Ardoz, Madrid, Spain
}

\begin{document}
%
%
%
%


\def\aj{\rm{AJ}}                   
\def\araa{\rm{ARA\&A}}             
\def\apj{\rm{ApJ}}                 
\def\apjl{\rm{ApJ}}                
\def\apjs{\rm{ApJS}}               
\def\ao{\rm{Appl.~Opt.}}           
\def\apss{\rm{Ap\&SS}}             
\def\aap{\rm{A\&A}}                
\def\aapr{\rm{A\&A~Rev.}}          
\def\aaps{\rm{A\&AS}}              
\def\azh{\rm{AZh}}                 
\def\baas{\rm{BAAS}}               
\def\jrasc{\rm{JRASC}}             
\def\memras{\rm{MmRAS}}            
\def\mnras{\rm{MNRAS}}             
\def\pra{\rm{Phys.~Rev.~A}}        
\def\prb{\rm{Phys.~Rev.~B}}        
\def\prc{\rm{Phys.~Rev.~C}}        
\def\prd{\rm{Phys.~Rev.~D}}        
\def\pre{\rm{Phys.~Rev.~E}}        
\def\prl{\rm{Phys.~Rev.~Lett.}}    
\def\pasp{\rm{PASP}}               
\def\pasj{\rm{PASJ}}               
\def\qjras{\rm{QJRAS}}             
\def\skytel{\rm{S\&T}}             
\def\solphys{\rm{Sol.~Phys.}}      
\def\sovast{\rm{Soviet~Ast.}}      
\def\ssr{\rm{Space~Sci.~Rev.}}     
\def\zap{\rm{ZAp}}                 
\def\nat{\rm{Nature}}              
\def\iaucirc{\rm{IAU~Circ.}}       
\def\aplett{\rm{Astrophys.~Lett.}} 
\def\apspr{\rm{Astrophys.~Space~Phys.~Res.}}
\def\bain{\rm{Bull.~Astron.~Inst.~Netherlands}} 
\def\fcp{\rm{Fund.~Cosmic~Phys.}}  
\def\gca{\rm{Geochim.~Cosmochim.~Acta}}   
\def\grl{\rm{Geophys.~Res.~Lett.}} 
\def\jcp{\rm{J.~Chem.~Phys.}}      
\def\jgr{\rm{J.~Geophys.~Res.}}    
\def\jqsrt{\rm{J.~Quant.~Spec.~Radiat.~Transf.}}
\def\memsai{\rm{Mem.~Soc.~Astron.~Italiana}}
\def\nphysa{\rm{Nucl.~Phys.~A}}   
\def\physrep{\rm{Phys.~Rep.}}   
\def\physscr{\rm{Phys.~Scr}}   
\def\planss{\rm{Planet.~Space~Sci.}}   
\def\procspie{\rm{Proc.~SPIE}}   

\let\astap=\aap
\let\apjlett=\apjl
\let\apjsupp=\apjs
\let\applopt=\ao

\maketitle

\begin{abstract}
We present a new Y dwarf, WISE J030449.03-270508.3, confirmed from a candidate sample designed to pick out low temperature objects from the WISE database. The new object is typed Y0pec following a visual comparison with spectral standards, and lies at a likely distance of 10-17 pc. Its tangential velocity suggests thin disk membership, but it shows some spectral characteristics that suggest it may be metal-poor and/or older than previously identified Y0 dwarfs. Based on trends seen for warmer late type T dwarfs, the $Y$-band flux peak morphology is indicative of sub-solar metallicity, and the enhanced red wing of the $J$-band flux peak offers evidence for high gravity and/or low metallicity (with associated model trends suggesting an age closer to $\sim$10 Gyr and mass in the range 0.02-0.03 $M_{\odot}$). This object may thus be extending the population parameter-space of the known Y0 dwarfs.
\end{abstract}

\begin{keywords}
surveys - stars: low-mass, brown dwarfs
\end{keywords}

\section{Introduction}
\label{sec:intro}

The discovery of Y dwarfs \citep{cushing2011} has extended the $T_{\rm eff}$ parameter-space of brown dwarfs \citep{chabrier2000} below 500 K, with Y0 dwarfs generally 400-450 K \citep{dupuy2013}, Y1 dwarfs 350-400 K \citep[e.g.][]{kirkpatrick2013}, and the most extreme example potentially as cool as $\sim$250 K \citep{luhman2014}. WISE J1828+2650 was first identified as the archetypal Y dwarf \citep[$\ge$Y2;][]{cushing2011} due to its extremely red near-mid-infrared colours and its red $J-H$ colour (suggestive of a Wien tail flux collapse), though parallax measurements have subsequently shown that this object is warmer than the Y0 dwarfs, with its extreme nature presumably influenced by other properties \citep{dupuy2013}. A significant decrease in the width of the $J$-band flux peak \citep[compared with the T9 spectral standard UGPS 0722-05;][]{lucas2010,cushing2011}, has become the practical means for separating late-T and Y dwarfs. Though some Y dwarf spectra (with sufficient signal-to-noise) also show possible evidence of NH$_3$ in the $H$-band, and some show enhanced $Y$-band flux \citep[but not all; see fig 22 of][]{mace2013a}, believed to be due to reduced absorption in the KI wings as potassium starts to form KCl \citep[a trend that starts for the latest T dwarfs;][]{leggett2010,lodieu2013}.

At time of writing there are seventeen spectroscopically confirmed Y dwarfs \citep{cushing2011,kirkpatrick2011,kirkpatrick2012,liu2012,tinney2012,kirkpatrick2013,cushing2014a}, and three additional objects with similar $T_{\rm eff}$ (or lower) for which it has not been possible to measure spectra due to their faintness in the near-infrared \citep[WD 0806-661B, CFBDSIR J1458+1013B, and WISE J0855-0714;][]{luhman2011,liu2011,luhman2014}.

This new spectroscopic class was revealed with the advent of the Wide-field Infrared Survey Explorer \citep[WISE;][]{wright2010} observatory, whose sensitivity in the mid-infrared opened up this detection-space. Previous large-scale surveys had searched for brown dwarfs in the near-infrared and red-optical, uncovering field L and T dwarfs \citep{kirkpatrick2005} with the Two Micron All Sky Survey \citep[2MASS;][]{skrutskie2006} and the Sloan Digital Sky Survey \citep[SDSS;][]{york2000}, and then probing to the coolest T dwarfs \citep[e.g.][]{warren2007,burningham2008,delorme2008a,burningham2010,lucas2010} with the UKIRT Infrared Deep Sky Survey \citep[UKIDSS;][]{lawrence2007}, the Canada France Brown Dwarf Survey \citep[CFBDS;][]{delorme2008b}, and more recently the Visible and Infrared Survey Telescope for Astronomy \citep[VISTA;][]{mcmahon2013,lodieu2012}.

WISE surveyed the whole sky in four filters centred at 3.4 $\mu m$, 4.6 $\mu m$, 12 $\mu m$, and 22 $\mu m$ (W1, W2, W3 and W4 respectively). In particular W1 covers a deep CH$_4$ absorption band in the spectra of cool brown dwarfs with W2 covering a spectral region of relatively low opacity. This leads to very red $W1-W2$ colours for cool brown dwarfs \citep{mainzer2011a} and facilitates colour-based brown dwarf searches \citep[e.g.][]{kirkpatrick2012}. WISE also has a time-domain, with baselines ranging from $\sim$36 hours \citep[identifying e.g. near-earth objects or NEOs as part of the NEOWISE programme;][]{mainzer2011b} up to 6-12 months when the main cryogenic mission data is combined with data from the post-cryogenic survey phase (the AllWISE data release). This facilitates proper motion searches such as \citet{luhman2013} and \citet{kirkpatrick2014}. It also offers greater scope to assess the nature of WISE sources via an analysis of their multiple measurements and comparison with known populations. \citet{pinfield2014} followed this approach, using a control sample (from SDSS) of isolated point-like non-variable non-moving sources to guide the identification of brown dwarf candidates down to low signal-to-noise.

Detailed studies of the full spectral diversity of Y dwarfs are important for our understanding of ultra-cool atmosphere physics \citep[e.g.][]{burrows2003,baraffe2003,burrows2007,saumon2008,allard2011,burrows2011,morley2012,morley2014}, with implications for giant exoplanet research as well as brown dwarfs \citep[the brown dwarf exoplanet connection; e.g.][]{pinfield2013,beichman2014,faherty2013,marocco2014,liu2013,canty2013}. Given the widely varying Y dwarf spectral energy distributions \citep[evident not only for WISE J1828+2650, but amongst other Y dwarfs sharing very similar $T_{\rm eff}$;][]{dupuy2013}, there is great incentive to calibrate this diversity against physical properties.

This requires the identification of Y dwarfs across their full parameter-space, as well as benchmarks systems with independently constrained properties \citep[following on from e.g.][]{pinfield2006,burningham2009,dayjones2011,pinfield2012,gomes2013,zhang2013,deacon2014}. Such samples will include the most unusual (and informative) observational outliers, as well as offering a means to calibrate the spectral sensitivities directly. Growth and increased diversity in the known Y dwarf population is thus important not only to test formation models \citep[e.g.][]{stamatellos2011,bate2012} through population studies \citep[following e.g.][]{pinfield2008,kirkpatrick2012,burningham2013,dayjones2013}, but also to reveal the rare examples that will offer the strongest tests for atmosphere models.

In this paper we present a new Y dwarf, WISE J030449.03-270508.3 (hereafter WISE J0304-2705), confirmed from amongst the candidate sample identified by \citet{pinfield2014}. Section \ref{sec:id} summarises the WISE candidate selection method. Sections \ref{sec:nirphot} and \ref{sec:spec} describes our near-infrared photometric and spectroscopic follow-up, with section \ref{sec:pm} detailing our proper motion measurement. Section \ref{sec:disc} discusses the spectral classification of WISE J0304-2705, and estimates for distance and kinematics. In addition we discuss our interpretation of the $J$-band spectral morphology of WISE J0304-2705, and the resulting theoretical implications for its surface gravity, age and mass. Conclusions are presented in Section \ref{sec:conc}.

\section{Identification}
\label{sec:id}

WISE J0304-2705 was first identified as a candidate late-type object in \citet{pinfield2014}. The search method identified WISE All-Sky sources detected in the $W2$-band only, and probed down to low signal-to-noise levels ({\tt{w2snr}}$\ge$8), targeting objects with faint $W2$ magnitudes and red $W1-W2$ colour. Spurious sources were removed using database selection criteria defined through analysis of a control sample comprising isolated point-like non-variable non-moving sources from the SDSS. A basic summary of the selection and rejection criteria is given below \citep[for a more detailed description see][]{pinfield2014};

\begin{itemize}
\item{Select sources only detected in the $W2$ band with signal-to-noise ${\tt{w2snr}}\ge8$.}
\item{Require at least 8 individual exposures (in all bands; ${\tt{w\star m}\ge8}$) covering the sky position.}
\item{The line of sight extinction must be $Av<$0.8 to remove reddened contamination.}
\item{Reject non-point-like sources for which the reduced $\chi^2$ of the $W2$ profile fit photometry (${\tt{w2rchi2}}$) was $>$1.2.}
\item{Reject sources for which the scatter in the multiple measurement photometry was higher than expected from the integrated flux uncertainty, if $\log{({\tt{w2sigp1}}-{\tt{w2sigmpro}})}>1.3-1.38\log{({\tt{w2snr}})}$.}
\item{Reject sources for which the number of detections in the individual $W2$ frames was less than expected, if $({\tt{w2nm}}/{\tt{w2m}})<1.8\log{({\tt{w2snr}})}-1.7$.}
\item{Reject faint sources ($W2$ signal-to-noise from 8-10) within extended bright star halo regions, by comparing with 2MASS point-source-catalogue positions \citep[see figure 2 of][]{pinfield2014}.}
\item{Select final candidates by visual inspection, rejecting artefacts, resolved extended structures (e.g. nebulosity and galaxies), badly blended sources, and sources with visual $W1$, $W3$ or $W4$ detections.}
\end{itemize}

\begin{figure}
\begin{center}
\includegraphics[height=8.0cm, angle=0]{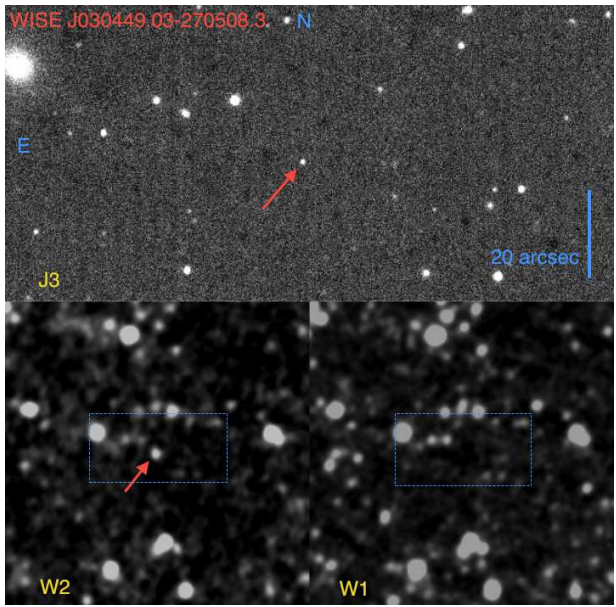}
\caption{Finder images of WISE J0304-2705 (indicated with a red arrow). The top image is from Four Star in the $J3$-band, and is 2$\times$1 arcmin. The lower two are WISE images in the $W2$- and $W1$-bands, and are 5$\times$5 arcmins on the side.  In the top image north and east are indicated in blue, and a 20 arcsecond scale bar is shown. In the WISE images the size of $J$-band image is indicated with a dashed line box.\label{fig:finder}}
\end{center}
\end{figure}

Figure \ref{fig:finder} shows WISE and follow-up images of WISE J0304-2705 (which is indicated with a red arrow). A $J$-band image is at the top (2$\times$1 arcmins), with the WISE $W2$ and $W1$ images below (each 5$\times$5 arcmins).

\section{Near-infrared follow-up}
\label{sec:nirphot}

WISE J0304-2705 was initially followed up using the SofI (Son of ISAAC) instrument mounted on the European Southern Observatory's New Technology Telescope on La Silla. Observations were obtained on 2013 January 1 under photometric conditions, seeing 0.8 arcsec, and dark conditions. SofI is equipped with a 1024$\times$1024 Hawaii HgCdTe offering a field-of-view of 4.9$\times$4.9 arcmin and a pixel size of 0.292 arcsec in its “Large-field” configuration. $J$-band images were measured using integrations of 10 seconds repeated 6 times with a 10-point dither pattern yielding a total exposure of 10 min. Dome flats were obtained during the afternoon preceding the observations. The images were reduced with the ESO SofI pipeline (via the {\tt{gasgano}} GUI). A faint low signal-to-noise source was seen in the resulting $J$-band image near the WISE position, so further follow-up was implemented.

Magellan / Four Star observations were obtained on 2013 August 15. Images were taken using the $J3$ and $J2$ filters. The $J2$ filter is centred at 1.144 microns (FWHM=0.136 microns), where methane absorption produces strong suppression for late-T and Y brown dwarfs. The $J3$ filter is centred at 1.288 microns (FWHM=0.140 microns), an opacity hole for cool brown dwarf photospheres \citep[see figure 1 of][]{tinney2012}. Observations used individual integrations of 20 seconds, with 6 co-adds and a random 11 point jitter pattern, leading to total exposure times of 22 mins. The target was centered on detector 2 as it has the most uniform pixel sensitivity, providing a 5.4 arcmin FOV. The data were reduced using standard IRAF routines. Calibration was achieved in two ways. For the $J3$ photometry 2MASS stars in the images were used \citep[with 2MASS to MKO conversions from][]{carpenter2001} to define the $J3$ zero points, following \citet{tinney2012}. Saturation limited the choice of calibrators, and we thus had to make use of a limited number of red calibrators \citep[foregoing the $0.4<(J-K)_{MKO}<0.8$ requirement of][]{tinney2012}. Our $J3$ zero point thus has sizable uncertainties at the level of $\sim$0.2 magnitudes. However, we were able to calibrate $(J3-J2)$ colour directly by measuring the instrumental colours of many sources in the $J3$ and $J2$ images, and calculating an offset that led to an average colour of $(J3-J2)$=0. This offset was then applied to the instrumental $(J3-J2)$ colour of WISE J0304-2705, which in turn was used to calibrate $J2$. We determined $(J3-J2)=-1.55\pm 0.1$, providing strong evidence for significant methane absorption \citep[which is known to produce colours of $J3-J2<-0.5$;][]{tinney2012}.

In addition, $J$ and $H$ photometry was obtained with FLAMINGOS-2 on Gemini South \citep{eikenberry2004} on 2013 December 22 (via program GS-2013B-Q-16), in the ``bright mode'' read mode. Nine 60-second images were obtained at $J$, and  thirty-six 10 second images at $H$, for total on-source exposures of 9 minutes and 6 minutes at $J$ and $H$ respectively. Routine photometric calibration is still being established for the instrument. We determined photometric zero-points using 15th to 16th magnitude 2MASS stars in the image, as well as observations made of photometric standards and of known late-type T dwarfs via a different Gemini program (Leggett personal communication 2014). These measurements show that the photometric system is equivalent to the Mauna Kea Observatories system \citep{tokunaga2005}, and that non-linearity is not significant over the range of data counts in our images. We adopted zero-points of 24.91$\pm$0.05 at $J$ and  25.15$\pm$0.05 at $H$, and used apertures of diameter $2\farcs9$ and $2\farcs2$ at $J$ and $H$ respectively, with annular sky regions and aperture corrections determined from stars in the image.

The photometric measurements of WISE J0304-2705, along with a summary of other measured and inferred properties, are presented in Table \ref{tab:properties}. The table gives WISE All-Sky and AllWISE magnitudes, with colours (here and hereafter) derived using AllWISE measurements.

\begin{table}
\begin{center}
\begin{tabular}{|l|r|}
\hline
\multicolumn{2}{|c|}{Measured and inferred properties of WISE J0304-2705} \\
\hline
R.A.(WISE) J2000 & 03h04m49.03s \\
Dec(WISE) J2000 & -27$^{\circ}$05'08.3'' \\
Spectral type & Y0pec \\
\hline
\multicolumn{2}{|c|}{Magnitudes} \\
$W2$(All-Sky) & 15.60$\pm$0.10 \\
$W1$(All-Sky) & $>$18.95 \\
$W2$(AllWISE) & 15.59$\pm$0.09 \\
$W1$(AllWISE) & $>$19.14 \\
$W3$(AllWISE) & $>$12.78 \\
$W4$(AllWISE) & $>$8.71 \\
$J_{MKO}$ & 20.79$\pm$0.09 \\
$H_{MKO}$ & 21.02$\pm$0.16 \\
$J3$ & 21.26$\pm$0.21 \\
$J2$ & 22.81$\pm$0.21 \\
\hline
\multicolumn{2}{|c|}{Colours (using AllWISE magnitudes)} \\
$W1-W2$ & $>$3.55 \\
$J_{MKO}-W2$ & 5.20$\pm$0.13 \\
$H_{MKO}-W2$ & 5.43$\pm$0.18 \\
$(J-H)_{MKO}$ & -0.23$\pm$0.18 \\
$J3-J2$ & -1.55$\pm$0.10 \\
\hline
\multicolumn{2}{|c|}{Proper motion (arcsec/yr)} \\
$\mu_{\alpha}\cos{\delta}$ & -0.03$\pm$0.10 \\
$\mu_{\delta}$ & 0.65$\pm$0.10 \\
$\mu_{tot}$ & 0.65$\pm$0.14 \\
H$_{W2}$ & 19.7$\pm$0.5 \\
\hline
\multicolumn{2}{|c|}{Distance and kinematics} \\
D & 10-17 pc \\
$V_{tan}$ & 24-64 km s$^{-1}$ \\
\hline
\end{tabular}
\end{center}
\caption{Coordinates, photometry and proper motion for WISE J0304-2705. \label{tab:properties}}
\end{table}

\section{Spectroscopy}
\label{sec:spec}

Long-slit spectroscopy was obtained with FLAMINGOS-2 on Gemini South on 2013 September 29 (via program GS-2013B-Q-16), in the ``faint mode'' read mode. Twenty-four 300 second images were obtained, for a total on-source time of two hours. The $JH$ grism and filter were used with the 4-pixel slit, producing a spectrum from 1.0 to 1.8 $\mu$m (see Figure \ref{fig:spec}). Due to a problem with the delivered image quality across the field of view, the resolution varied from 4 nm in the center of the spectral range, to 7 nm at the extremes of the range, for a resolving power of 140 to 300. This resolution is sufficient for the purposes of classifying the object and broadly assessing its spectral morphology. The F3V star HD 27975 was used as a calibrator to remove telluric features. Flat fields and spectral arcs were obtained using lamps in the calibration unit mounted on the telescope. The data were reduced in the standard way using IRAF routines, with flux calibration achieved through division by the IRTF Spex library spectrum of HD 26015 \citep{rayner2009}. Although HD 27975 and HD 26015 have B-V of 0.47 and 0.36 respectively (suggesting some extinction for HD 27975), we did not correct for this as it would not be significant above the level of the spectral signal-to-noise. The signal-to-noise in the peak of the $Y$-, $J$- and $H$-bands is 4, 5 and 3 respectively.

\section{Proper motion measurement}
\label{sec:pm}

We measured the proper motion of WISE J0304-2705 using our SofI $J$-band image as a first epoch (2013 January 1), and our FLAMINGOS-2 $J$ image as a second epoch (2013 December 22), leading to a baseline of 0.97 years. We selected 21 reference stars to define an inter-epoch positional transform, which had associated root-mean-square residuals (when compared to the reference stars) of 0.05 arcsecs. $J$-band counterpart centroiding uncertainty was estimated to be 0.08 arcsecs by assessing the scatter in centroid measurements of a simulated population of sources with Gaussian point spread functions and added Poisson noise. Fit residuals and centroiding uncertainties were combined in quadrature to give full proper motion uncertainties. WISE J0304-2705 was found to be a high proper motion source, moving at 0.65$\pm$0.14 arcsec/yr. The proper motion measurements are listed in Table \ref{tab:properties}.

\section{Discussion}
\label{sec:disc}

\subsection{Spectral classification}
\label{sec:class}

\begin{figure}
\begin{center}
\includegraphics[height=7.0cm, angle=0]{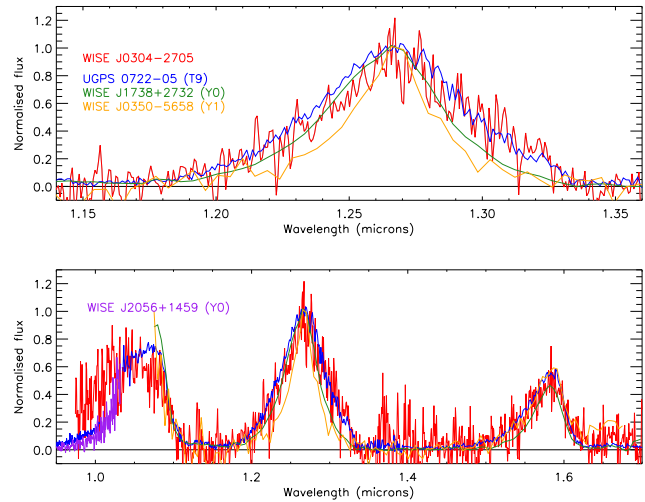}
\caption{The near-infrared Flamingos-2 spectrum of WISE J0304-2705 (plotted in red). The top plot shows the $J$-band region, and the lower plot the full YJH region. Several spectra are over-plotted for spectral typing and comparison; UGPS 0722-05 (T9 spectral standard in blue), WISE J1738+2732 (Y0 spectral standard in green), WISE J0350-5658 (Y1 spectral standard in orange), and red-optical WISE J2056+1459 (Y0 in purple). All spectra are normalised to an average of unity from 1.265-1.270 microns. \label{fig:spec}}
\end{center}
\end{figure}

The top plot in Figure \ref{fig:spec} shows a zoom-in of the $J$-band flux peak region, with the full $YJH$ spectrum in the lower plot. The spectrum of WISE J0304-2705 is shown in red. For visual spectral typing and comparison we over-plot the spectral standards UGPS 0722-05 (T9; blue), WISE J1738+2732 (Y0; green), and WISE J0350-5658 (Y1; orange), as well as the red-optical spectrum of the Y dwarf WISE J2056+1459 \citep{cushing2011,leggett2013}. The spectra have been normalised to an average of unity in the 1.265-1.270 micron range (including WISE J2056+1459, though near-infrared not shown).

The highest signal-to-noise is in the central region of the $J$-band flux peak, and from 1.22-1.29 microns the spectrum of WISE J0304-2705 is well represented by the Y0 spectral standard. It is a poor match to the T9 standard in this range, and at shorter $J$-band wavelengths the noise precludes a useful comparison. However, over the 1.29-1.33 micron range the average flux level of WISE J0304-2705 is brighter than the Y dwarf standards, being closer to the T9 standard. In addition, in the $Y$-band (over at least the 0.974-1.033 micron range covered by the spectra) it can be seen that WISE J0304-2705 is relatively bright compared to the comparison Y0 dwarf WISE J2056+1459. Aside from these two main differences (which will be discussed further in Section \ref{sec:unusual}) the full spectrum of WISE J0304-2705 is well represented by the Y0 spectral standard. To assess the $J$-band flux enhancement we define a new spectral ratio J-wing as follows.

\begin{equation}
{\rm J-wing}=\frac{\tilde{F}_{1.290-1.330\mu m}}{\tilde{F}_{1.265-1.270\mu m}}
\end{equation}

The denominator wavelength range is the same as for the J-narrow ratio, but the numerator fully samples the enhanced flux region for WISE J0304-2705. We note that the CH4-J ratio \citep{burgasser2006} also samples this region at some level, but only $>$1.315 microns. Table \ref{tab:ratios} presents spectral ratio measurements for WISE J0304-2705 as well as for the T9, Y0 and Y1 spectral standards \citep[median ratios were derived following][]{mace2013a}. Spectral type average ratios are also listed aside the T9 and Y0 spectral standard values \citep[from][]{mace2013a}. We measured $J$-band ratios (including previously defined ratios and the new J-wing ratio), as well as flux peak ratios involving the $Y$, $J$ and $H$ flux peaks. We do not present $H$-band ratios since the signal-to-noise of the WISE J0304-2705 spectrum is too low for these to be useful.

\begin{table}
\begin{center}
\begin{tabular}{|l|c|c|}
\hline
\multicolumn{3}{|l|}{\bf WISE J0304-2705} \\
Ratio       & Median           & Integrated \\
\hline
H2O-J       & -0.009$\pm$0.037 & -0.016$\pm$0.026 \\
CH4-J       &  0.150$\pm$0.031 &  0.167$\pm$0.021 \\
W$_J$       &  0.181$\pm$0.025 &  0.200$\pm$0.018 \\
J-narrow    &  0.759$\pm$0.054 &  0.776$\pm$0.040 \\
Y/J         &  0.733$\pm$0.052 &  0.702$\pm$0.035 \\
H/J         &  0.538$\pm$0.034 &  0.532$\pm$0.023 \\
J-wing      &  0.389$\pm$0.039 &  0.409$\pm$0.026 \\
\hline
\multicolumn{3}{|l|}{\bf Spectral type comparisons} \\
Ratio       & Spectral standard$^1$ & Average (per half type)$^2$ \\
\hline
\multicolumn{3}{|c|}{\bf T9} \\
H2O-J       &  0.041$\pm$0.027$^2$ &  0.032$\pm$0.037 \\
CH4-J       &  0.156$\pm$0.026$^2$ &  0.117$\pm$0.030 \\
W$_J$       &  0.204$\pm$0.052$^2$ &  0.203$\pm$0.038 \\
J-narrow    &  0.862$\pm$0.040$^2$ &  0.879$\pm$0.053 \\
Y/J         &  0.420$\pm$0.050$^2$ &  0.448$\pm$0.095 \\
H/J         &  0.530$\pm$0.020$^2$ &  0.555$\pm$0.030 \\
J-wing      &  0.320$\pm$0.005     &  - \\
\hline
\multicolumn{3}{|c|}{\bf Y0} \\
H2O-J       &  0.039$\pm$0.009$^2$ & -0.022$\pm$0.050 \\
CH4-J       &  0.052$\pm$0.010$^2$ &  0.045$\pm$0.031 \\
W$_J$       &  0.122$\pm$0.028$^2$ &  0.117$\pm$0.043 \\
J-narrow    &  0.839$\pm$0.027$^2$ &  0.778$\pm$0.050 \\
Y/J         &  -                   &  0.423$\pm$0.131 \\
H/J         &  0.460$\pm$0.020$^2$ &  0.467$\pm$0.061 \\
J-wing      &  0.164$\pm$0.012     &  - \\
\hline
\multicolumn{3}{|c|}{\bf Y1} \\
H2O-J       & -0.055$\pm$0.071$^2$ &  - \\
CH4-J       &  0.068$\pm$0.049$^2$ &  - \\
W$_J$       &  0.116$\pm$0.096$^2$ &  - \\
J-narrow    &  0.676$\pm$0.095$^2$ &  - \\
Y/J         &  -                   &  - \\
H/J         &  0.751$\pm$0.081     &  - \\
J-wing      &  0.136$\pm$0.038     &  - \\
\hline
\multicolumn{3}{|l|}{$^1$ T9=UGPS 0722-05, Y0=WISE J1738+2732} \\
\multicolumn{3}{|l|}{Y1=WISE J0350-5658} \\
\multicolumn{3}{|l|}{$^2$ Taken from \citet{mace2013a}.} \\
\hline
\end{tabular}
\end{center}
\caption{Spectral ratios for WISE J0304-2705, and the T9, Y0 and Y1 comparison objects. Median flux ratios were calculated using multiple realizations of a Monte Carlo spectrum model with flux density values (at each wavelength) randomly drawn from a normal distribution with mean and standard deviation set by the measured values and associated uncertainties.\label{tab:ratios}}
\end{table}

In the $J$-band peak, the J-narrow ratio value is mid-way between the Y0 and Y1 spectral standard values. The Wj ratio value is similar to T9 expectations, but the uncertainties are relatively large due to the low signal-to-noise in the numerator region, and this ratio does not strongly differentiate between T9 and Y. The value of J-wing for WISE J0304-2705 is comparable to expectations for T9, being a factor $\sim$2 greater than for the Y0 spectral standard. Also, the value of CH4-J is similar to T9, being rather greater than the range expected for Y dwarfs \citep[with a previously observed maximum CH4-J for Y0 dwarfs of 0.096$\pm$0.010 for WISE J2056+1459; see][]{mace2013a}. The higher values for J-wing and CH4-J result from the relatively enhanced flux in the red wing of the $J$-band flux peak. The H2O-J ratio is saturated, and the measured value does not differentiate between late-T and Y types. Considering the full $YJH$ spectrum it can be seen that the Y/J ratio is significantly greater than expectations for either T9 or Y0 spectral types. A $Y$-band enhacement is also clearly seen by direct comparison with the spectrum of WISE J2056+1459. The H/J ratio is very similar to the Y0 spectral standard and Y0 average value, being rather lower than the T9 and Y1 values.

We base our spectral classification of WISE J0304-2705 primarilly on the visual comparison with the spectral standards in the $J$-band wavelength range 1.22-1.29 microns. This region offers the best signal-to-noise, and avoids the 1.29-1.33 micron range where we note relative enhancement (as will be discussed in Section \ref{sec:unusual}, models suggest this region is sensitive to gravity and metallicity, and as such is not ideal for spectral classification). Our typing also acknowledges the unusual $J$-band morphology that makes the object and outlier in e.g. J-narrow J-wing space. We thus classify WISE J0304-2705 as Y0pec. Only one other Y0 dwarf has been classified as peculiar (WISE J1405+5334), which was as a result of its $H$-band flux peaking at a wavelength 60\AA redder compared to other Y dwarfs and the T9 spectral standard.

\subsection{Distance, kinematics and reduced proper motion}
\label{sec:distance}

In Table \ref{tab:distances} we present spectrophotometric distances estimates for WISE J0304-2705, derived using the available $J$-, $H$- and $W2$-band photometry. We did the calculations assuming a spectral type of Y0, and defined absolute magnitudes using \citet{dupuy2013}, converting from $M_{[4.5]}$ into $M_{W2}$ using an average colour of $W2-[4.5]$=0.08$\pm$0.06 (for the 3 Y0 dwarfs used in their calculation).

\begin{table}
\begin{center}
\begin{tabular}{|l|c|}
\hline
     & Distance (pc) \\
\hline
$J_{MKO}$ & 12.2-15.6 \\
$H_{MKO}$ & 10.4-14.1 \\
$W2$      & 12.9-16.9 \\
\hline
\end{tabular}
\end{center}
\caption{Spectrophotometric distance estimates for WISE J0304-2705. Estimates are made using the $J$, $H$ and $W2$ apparent magnitudes, and assuming a Y0 spectral type. \label{tab:distances}}
\end{table}

\begin{figure*}
\begin{center}
\includegraphics[width=15.0cm, angle=0]{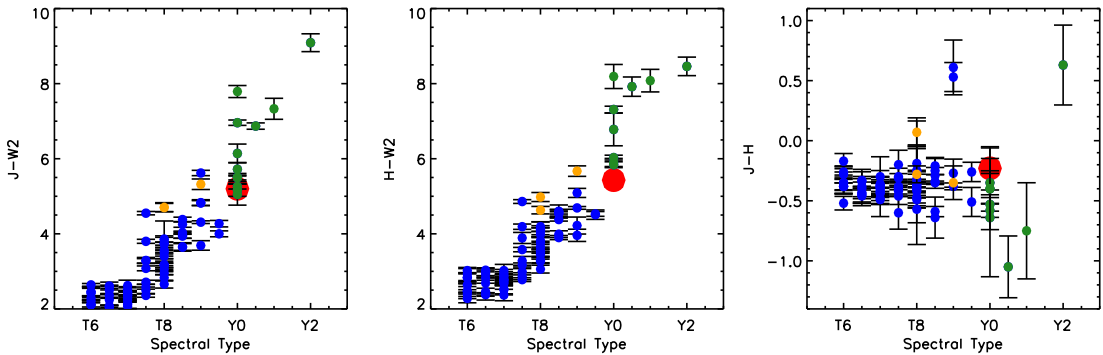}
\caption{Colour versus spectral type plots. WISE J0304-2705 is plotted in red. L and T dwarfs from \citet{leggett2013} are plotted in blue (ignoring known unresolved binaries). Late-T subdwarfs with thick-disk/halo kinematics are shown in orange. Available Y dwarf photometry is plotted in green.\label{fig:col_spt}}
\end{center}
\end{figure*}

The three bands yield distance constraints that overlap, covering a range 10.4-16.9 pc. This is not surprising when one considers Figure \ref{fig:col_spt}. These plots place WISE J0304-2705 in the colour-spectral-type space of the late-T and Y dwarf population \citep[data from][]{leggett2013,mace2013b,pinfield2014,kirkpatrick2014,burningham2014}, and it can be seen that WISE J0304-2705 occupies essentially the same colour space as the previously known Y0 dwarfs, though at the bluer end in $J-W2$ and $H-W2$. Our best estimate for the distance of WISE J0304-2705 is thus 10-17 pc, appropriate for a single object. However, if WISE J0304-2705 was found to be an unresolved binary then it could be as distant as 24 pc.

\begin{figure*}
\begin{center}
\includegraphics[width=15.0cm, angle=0]{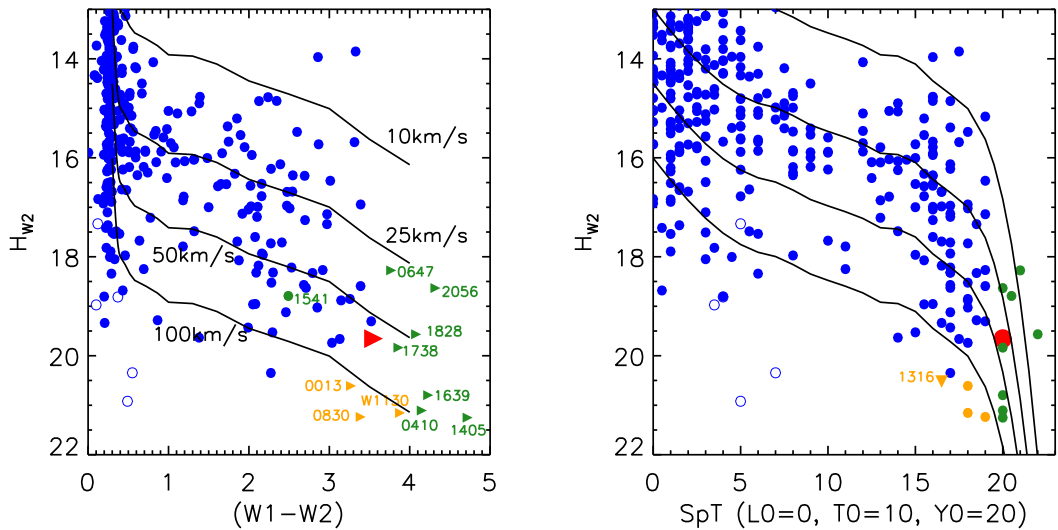}
\caption{Reduced proper motion (H$_W2$) plotted against $W1-W2$ colour (left hand plot) and spectral type (right hand plot). WISE J0304-2705 is plotted in red. L and T dwarfs (from the WISE team's census and other WISE cross-matches from the literature) are plotted as blue circles (subdwarfs are open symbols). Y dwarfs with measured proper motion are plotted in green. Four late-T sub-dwarfs with thick-disk/halo kinematics are plotted in orange (WISE J0013+0634; WISE J0833+0052; Wolf 1130B; ULAS J1316+0755). Right- and down-facing triangles indicate lower limits. Tracks of constant tangential velocity are shown in black. \label{fig:rpmd}}
\end{center}
\end{figure*}

By combining our distance constraint with the measured proper motion of WISE J0304-2705, we estimate a tangential velocity of $V_{tan}$=24-64 km~s$^{-1}$ which is consistent with thin disk membership \citep[e.g.][]{nissen2004,faherty2009}. In addition Figure \ref{fig:rpmd} plots WISE J0304-2705 in reduced proper motion diagrams (as a filled red symbol). Reduced proper motion ($H_{W2}=W2+5\log{\mu}+5$) is shown against both $W1-W2$ colour and spectral type, with disk L and T dwarfs shown as filled blue circles, and other Y dwarfs in green. Sub dwarfs from the thick-disk/halo are blue open circles (L dwarfs) or orange symbols \citep[T dwarfs;][]{mace2013b,pinfield2014,burningham2014}. Black lines delineate $V_{tan}$ ranges within the diagrams, and WISE J0304-2705 can be seen to lie in the middle of the Y dwarf population. Kinematically, WISE J0304-2705 thus appears to be a typical thin disk Y dwarf.

\subsection{Unusual spectral features}
\label{sec:unusual}

The full spectrum of WISE J0304-2705 (see Figure \ref{fig:spec}) shows that the $Y$-band flux peak is unusual. It is significantly broadened to the blue, with a peak between $\sim$1.02-1.04 microns. By contrast the $Y$-band spectrum of the comparison Y0 dwarf WISE J2056+1459 (in purple) is fainter and rising steeply over this range, with its $Y$-band flux peaking closer to the typical value for most late-T dwarfs ($\sim$1.08 microns). These unusual $Y$-band characteristics have been considered likely indicators of sub-solar metallicity for some time \citep{burrows2002,burgasser2006b}, and have recently been observed in the spectrum of the benchmark T8 Wolf 1130B, a companion to an M subdwarf with metallicity [Fe/H]=-0.64$\pm$0.17 \citep{mace2013b}. By contrast the mildly metal-poor ([Fe/H]=-0.38$\pm$0.06) T8 benchmark BD+01~2920 has a $Y$-band flux peak that is very similar to the T8 spectral standard 2MASS J0415-09 \citep[see fig 2 of][]{pinfield2012}. Assuming this metallicity sensitivity extends over the T/Y transition, then WISE J0304-2705 may have lower metallicity than the Y0 dwarf WISE J2056+1459, and could have sub-solar composition.

WISE J0304-2705 also shows a relative flux enhancement in the $J$-band (over the 1.29-1.33 micron range), when compared to the Y spectral standards. To assess if unresolved binarity might lead to such spectral morphology we considered a T9+Y0 composite spectrum. Two such binaries were studied by \citet{liu2012}, with the components of WISE J1217+1626AB having a $J$-band brightness difference of 2.10 magnitudes and the components of CFBDSIR J1458+1013AB a difference of 2.02 magnitudes. The average $J$-band brightness difference between single T9 and Y0 dwarfs is $\sim$1.7 magnitudes \citep{dupuy2013}. We compared the spectrum of WISE J0304-2705 to composite T9+Y0 spectra (constructed using the spectral standard spectra) covering a range of possible $J$-band magnitude differences from 1.5-2.0 magnitudes. None of the composite spectra came close to replicating the $J$-band morphology of WISE J0304-2705.

For a single object explanation of the $J$-band flux enhancement comparison with T8 benchmarks is once again instructive, as the red wing of the $J$-band flux peak of Wolf 1130B also shows an unusual morphology. Figure 5 of \citet{mace2013b} shows that this object has greatly enhanced flux in the red wing of the $J$-band flux peak compared to BD+01~2920 (which has a similar $J$-band flux peak to the T8 standard 2MASS J0415-09). This type of spectral morphology is thus characteristic of very late objects that are old (with relatively high surface gravity) and/or have low metallicity, although the enhancement seen for WISE J0304-2705 is at a significantly lower level than for Wolf 1130B.

\begin{figure*}
\begin{center}
\includegraphics[width=15.0cm, angle=0]{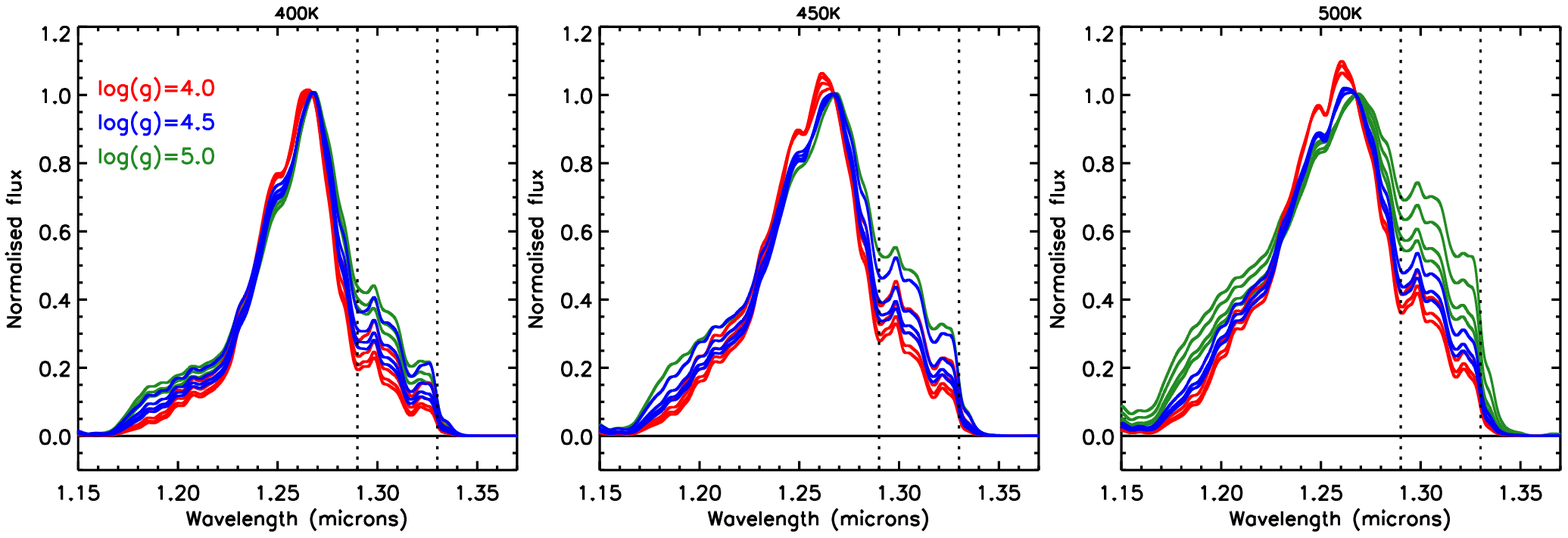}
\caption{Plots of the $J$-band spectral region showing the \citet{morley2012} model spectra. The three plots show model $T_{\rm eff}$=400K, 450K and 500K respectively. Different surface gravities ($\log{g}$=4.0, 4.5 and 5.0) are plotted in red, blue and green respectively. For each value of $\log{g}$ there are up to four spectra representing different sedimentation efficiencies ($f_{sed}$=2, 3, 4, 5). Vertical dotted lines enclose the spectral range 1.29-1.33 microns. \label{fig:cool_models}}
\end{center}
\end{figure*}

We have analysed this spectroscopic feature in the context of $\log{g}$ sensitivity, through comparison with the cloudy models of \citet{morley2012}. In Figure \ref{fig:cool_models} model spectra are shown for $T_{\rm eff}$ values of 500K, 450K and 400K \citep[covering the Y0 $T_{\rm eff}$ range in e.g.][]{dupuy2013}, and $\log{g}$ of 4.0, 4.5 and 5.0 (red, blue and green respectively) appropriate for the $T_{\rm eff}$ range. For each $T_{\rm eff}$ and $\log{g}$ combination the models are available with up to four different sedimentation efficiencies ($f{sed}$=2, 3, 4, 5). The 1.29-1.33 micron range is enclosed by dashed lines in Figure \ref{fig:cool_models}, and it can be seen that the models have significant sensitivity to $\log{g}$ in this range.

We can use the calculated trends with gravity for the J-wing spectral ratio to estimate gravity and hence age for WISE J0304-2705. Table \ref{tab:ratios} shows that the value of this spectral ratio is a factor of at least two larger than the equivalent value for the Y0 standard (Table \ref{tab:enhancements}). Using the model suite described in the previous paragraph, we calculated J-wing ratios as a function of gravity for each set of [Teff,fsed]. Table \ref{tab:enhancements} gives the resulting enhancement as gravity is increased (where an enhancement factor of 1 means no enhancement).

\begin{table}
\begin{center}
\begin{tabular}{|c|c|}
\hline
Ratio / Ratio & Enhancement \\
              & factor      \\
\hline
J-wing (0304) / J-wing (Y0)$^1$ & 1.99-2.82 \\
\hline
$\Delta\log{g}$ & Enhancement \\
                & factor      \\
\hline
4.0$\rightarrow$4.5 & 1.18-1.41 \\
4.5$\rightarrow$5.0 & 1.30-1.46 \\
4.0$\rightarrow$5.0 & 1.56-2.06 \\
\hline
\multicolumn{2}{|l|}{$^1$Y0=WISE J1738+2732} \\
\hline
\end{tabular}
\end{center}
\caption{Enhancement levels in the 1.29-1.33 micron range, as represented by the J-wing ratio. In the top of the table the relative enhancement for WISE J0304-2705 is given relative to the Y0 spectral standard. The lower portion of the table presents theoretical predictions for $T_{\rm eff}$=400-500K, $f_{sed}$=2,3,4,5. \label{tab:enhancements}}
\end{table}

For the theoretical $\log{g}$ changes shown in Table \ref{tab:enhancements} the only one that yields enhancement consistent with WISE J0304-2705 is $\log{g}$=4.0$\rightarrow$5.0, which would suggest an age close to 10 Gyr and a mass of 0.02-0.03$M_{\odot}$ \citep{saumon2008}. However, a more complete theoretical analysis should also assess the effects of sub-solar metallicity as well as high surface gravity, which requires the generation of new model spectra.

\begin{figure*}
\begin{center}
\includegraphics[width=13.0cm, angle=0]{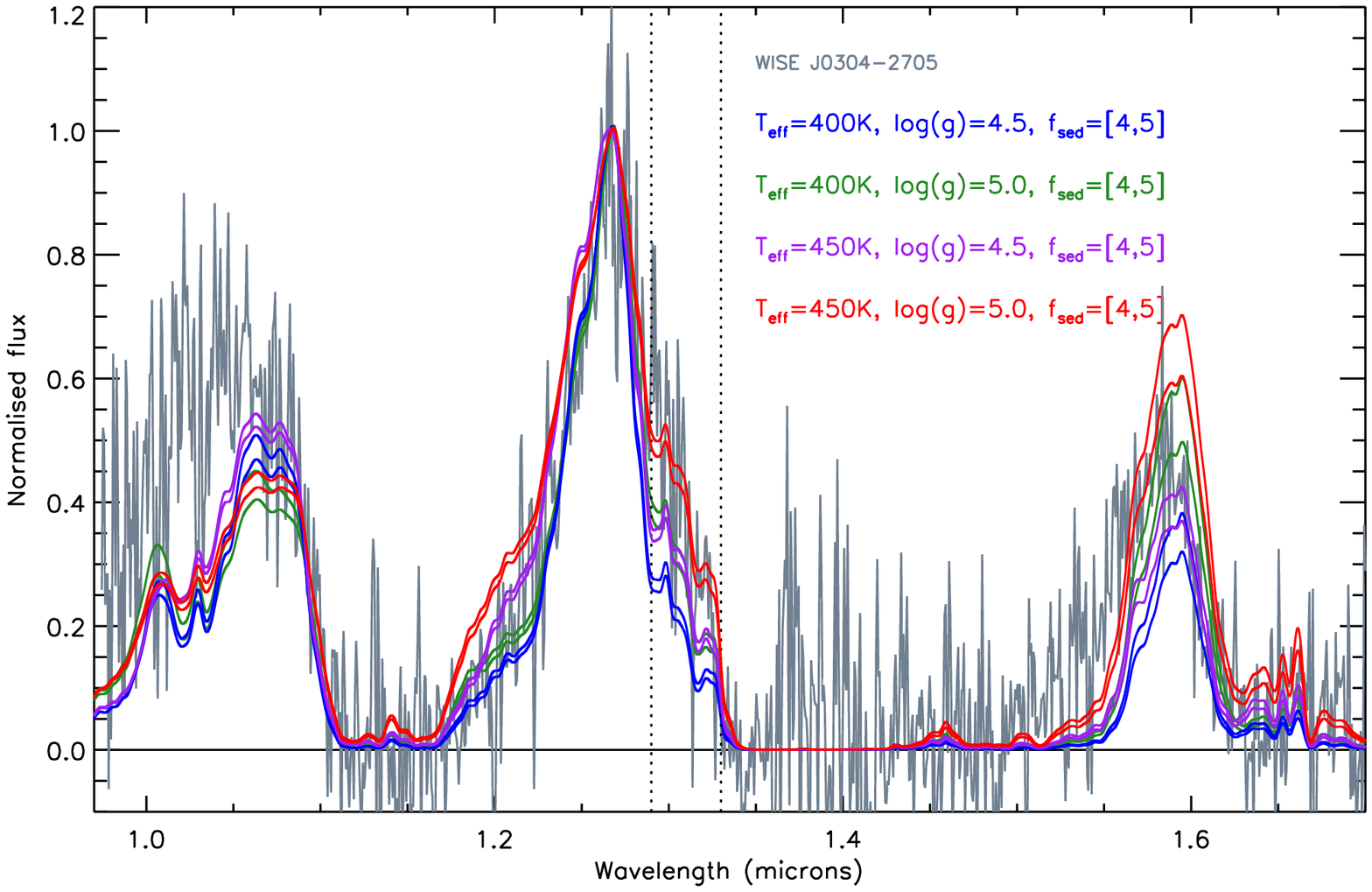}
\caption{The spectrum of WISE J0304-2705 with models over-plotted, as indicated. All spectra have been normalised to an average of unity from 1.265-1.270 microns. \label{fig:overplot_models}}
\end{center}
\end{figure*}

We have also made a direct comparison with the model spectra. Figure \ref{fig:overplot_models} shows the $YJH$ spectrum of WISE J0304-2705 (grey) with the indicated models over-plotted ($T_{\rm eff}$=400 and 450 K, $\log{g}$=4.5 and 5.0, and $f_{sed}$=4 and 5). These models offer a reasonable matched with the observed relative brightness of the $J$- and $H$-band flux peaks. The observed $Y$-band flux peak shows significantly different morphology than the models, as expected from our previous discussion. In the $J$-band the 400 K models give the best match to the blue wing of the flux peak, with the 450 K models being relatively brighter. In the red wing of the flux peak the gravity sensitive enhancement is apparent between the $\log{g}$=4.5 and 5.0 models, though the high gravity 450 K model gives the closest match.

\section{Conclusions and future work}
\label{sec:conc}

The discovery of WISE J0304-2705 identifies the 18th spectroscopically confirmed Y dwarf, which includes 17 that were identified as WISE sources and one that was a component of an unresolved WISE multiple system. A comparison between the $J$-band spectrum of WISE J0304-2705 and previously known Y dwarfs shows that this object has some unusual spectral features, including relatively enhanced flux in the $J$- (1.29-1.33 microns) and $Y$- (inclusive of 0.974-1.033 microns) bands. Based on observed trends for the warmer late-type T dwarfs these enhancements are both characteristic of sub-solar metallicity objects that may be old and high gravity.

In the future, additional higher signal-to-noise spectroscopy of WISE J0304-2705 can assess the stability of the relative flux enhancements, constraining any variability in these features \citep[i.e. if Y dwarf weather is important;][]{cushing2014b}. Adaptive optics imaging will be useful to assess if WISE J0304-2705 is an unresolved multiple or a single object, and thus if its spectrum may be a composite of more than one spectral type. Radial velocity measurements, as a further test of multiplicity and to determine space motion, are currently beyond the capabilities of the largest telescopes. But a parallax measurement combined with Spitzer photometry will be crucial for determining a robust $T_{\rm eff}$ constraint, and $K$-band photometry of WISE J0304-2705 (and other Y dwarfs) will show if this object is $K$-band suppressed, a known signature of high surface gravity and/or low metallicity for T dwarfs \citep[e.g.][]{pinfield2012}, caused by increased collisionally induced H$_2$ absorption.

We did not find any common proper motion companions to WISE J0304-2705 out to a separation of 10 arcmins, using the Hipparcos catalogue \citep{vanleeuwen2007} and the SuperCOSMOS Science Archive \citep{hambly2001}. Also, we found no kinematic membership of the moving groups listed in \citet{clarke2010} and \citet{torres2008} after assessing proper motion and distance constraints for a possible range of radial velocity (-200 km~s$^{-1}$ to +200 km~s$^{-1}$). However, as the Y dwarf population grows, benchmark systems should become more numerous. We note that WISE J0647-6232 \citep[Y1:][]{kirkpatrick2014} could be a benchmark Y dwarf in the Columba moving group, though radial velocity confirmation is not currently possible. And WD 0806-661B is a benchmark object but does not have a measured spectrum. As more Y dwarfs are uncovered, the confluence of observed spectral variations with predictions from theory and calibration from benchmark systems, should lead to an improved understanding of brown dwarfs and giant exoplanets with $T_{\rm eff}<$500 K.

\section*{Acknowledgments}

This publication makes use of data products from the Wide-field Infrared Survey Explorer, which is a 
joint project of the University of California, Los Angeles, and the Jet Propulsion Laboratory/California 
Institute of Technology, funded by the National Aeronautics and Space Administration.
This paper includes data gathered with the 6.5 meter Magellan Telescopes located at Las Campanas Observatory, Chile.
Based on observations obtained at the Gemini Observatory, which is operated by the Association of Universities 
for Research in Astronomy, Inc., under a cooperative agreement with the NSF on behalf of the Gemini partnership: 
the National Science Foundation (United States), the National Research Council (Canada), CONICYT (Chile), the 
Australian Research Council (Australia), Ministério da Ci\^{e}ncia, Tecnologia e Inova\c{c}\~{a}o (Brazil) and Ministerio de 
Ciencia, Tecnolog\'{i}a e Innovaci\'{o}n Productiva (Argentina).
Based on observations made with ESO Telescopes at the La Silla Paranal Observatory under programme ID ppp.c-nnnn.
DP, NL and ADJ have received support from RoPACS during this research and JG was supported by 
RoPACS, a Marie Curie Initial Training Network funded by the European Commission’s Seventh Framework Programme.
BB is supported by a Science and Technology Research Council (STFC) Consolidated Grant.
MG is financed by the GEMINI-CONICYT Fund, allocated to the project 32110014.
SKL is supported by the Gemini Observatory, which is operated by AURA, on behalf of the international 
Gemini partnership of Argentina, Australia, Brazil, Canada, Chile, the United Kingdom, and the United 
States of America.
NL is a Ramon y Cajal at the IAC in Tenerife (fellowship number 08-303-01-02) and is funded by the national programme AYA2010-19136 of the Spanish ministry of economy and competitiveness (MINECO).
RK acknowledges partial support from FONDECYT through grant 1130140.
ADJ was supported by a Fondecyt Postdoctorado under project number 3100098.
MTR received support from PB06 (CONICYT).
This research has made use of the SIMBAD database, operated at CDS, Strasbourg, France.

\bibliographystyle{mn2e}
\bibliography{refs}

\end{document}